\documentclass[12pt]{article}

\input epsf
\def\beq{\begin{eqnarray}}
\def\eeq{\end{eqnarray}}

\def\lsim{\mathrel{\rlap{\lower3pt\hbox{\hskip0pt$\sim$}}
    \raise1pt\hbox{$<$}}}         
\def\gsim{\mathrel{\rlap{\lower4pt\hbox{\hskip1pt$\sim$}}
    \raise1pt\hbox{$>$}}}         

\begin{document}


\vskip 1cm
\begin{center}

{\Large \bf Infrared Modification of Gravity$^{*}$}

\vskip 1cm {Gia Dvali}

\vskip 1cm
{\it Center for Cosmology and Particle Physics, Department of Physics, New York University, New York, NY 10003}\\
\end{center}

\vspace{0.9cm}
\begin{center}
{\bf Abstract}
\end{center}

 In this lecture I address the issue of possible large distance modification of gravity and its
observational consequences. Although, for the illustrative purposes we focus on a particular simple 
generally-covariant example, our conclusions are rather general and apply to large class of 
theories in which, already at the Newtonian level, gravity changes the regime at a certain very large
crossover distance $r_c$.  In such theories the cosmological evolution gets dramatically  modified
at the crossover scale, usually exhibiting a "self-accelerated" expansion, which can be 
differentiated from more conventional "dark energy" scenarios by precision cosmology.  
However, unlike the latter scenarios, theories of modified-gravity are extremely constrained 
(and potentially testable) by the precision gravitational measurements at much shorter scales. 
The reason is that modification implies the new propagating light degrees of freedom 
(additional polarizations of graviton) which penetrate at short distances in a rather profound way, and 
lead to the deviations from Einstein's gravity at the source-dependent scales. 
Despite the presence of extra polarizations of graviton, the theory is compatible with observations, 
since the naive perturbative  expansion in Newton's constant breaks down
at a certain intermediate scale.  This happens because the extra polarizations have couplings
singular in $1/r_c$. However, the correctly resummed non-linear solutions are 
regular and exhibit continuous Einsteinian limit.  Contrary to the naive expectation,  explicit examples indicate
that the resummed solutions remain valid after the 
ultraviolet completion of the theory, with the loop corrections taken into account.

\vspace{4cm}
$^{*)}$ Based on the talk given at Nobel Symposium on Cosmology and String
Theory, August 03, Sigtuna, Sweden.

\vspace{0.1in}

\newpage

\section{Introduction and Framework.}
 
 The question that  I will try to address in this talk is whether it is possible to self-consistently
modify General Relativity (GR) in far infrared (IR), and whether such a modification could come from some underlying fundamental theory.  Under the self-consistent  modification I mean modification that is
generally-covariant, and is free of unphysical negative or complex norm states (i.e., is ghost-free). 
I shall only be interested in theories that are not reducible to a simple addition
(to the ordinary gravity) of new light scalar or vector degrees of freedom. 

 Such theories are motivated by the "Dark Energy" and the "Cosmological Constant"  problems. However, even if one still hopes to find a less radical solution to these problems, the question of self-consistent  IR-modification of gravity is a valid fundamental question.  

 As we shall see, the answer to this question is probably "yes", as there is at least one existence
proof for such modifications (and probably there are many others).  However, the theories in question
are extremely constrained, because any modification of gravity (respecting locality and causality)
in arbitrary far IR implies existence of
new gravitational degrees of freedom (new polarizations of graviton) that inevitably penetrate
at short distances and lead to the potentially-observable effects. 

 If there is some new gravitational dynamics in far IR, in an effective short-distance action 
it would manifest itself in form of some effectively-non-local operators (possibly the infinite series) 
that should dominate over the standard Einstein-Hilbert action at large distances,  or equivalently 
at small curvatures.
Such operators should be suppressed by a crossover scale ($r_c$) beyond which the new dynamics takes over and gravity changes the regime. 
 \begin{equation}
\label{R+}
\int d^4x \sqrt{-g} \, M_P^2 \,  R \, + \,  [{\rm something~dominating~at} ~r > r_c]\, . 
\end{equation}
One could naively think that at short distances $r \ll r_c$, the effect of the new terms can be made arbitrarily small. However, this is not the case. General Relativity is an extremely constrained theory.  
Requirement of the massless graviton plus general-covariance uniquely fixes  the low energy action to
be the Einstein-Hilbert action. Hence, any deviation from it implies that gravity can no longer be
mediated by a massless spin-2 particle, but should inevitably include extra states.

 I shall discuss the above features on an example of a simple generally-covariant theory that modifies
gravity at large distances, however, our qualitative conclusions  will be very general and 
apply to large class of IR-modified theories (at least the ones that at the linearized level admit
a sensible  spectral representation).   The two emerging features of such IR-modified theories
are the (expected) strong modification of the cosmological evolution at the scale $r_c$, and also a
strong modification of laws of gravity at  a {\it source-dependent}  scale $r_*$.
\footnote{In a sense this is one and the same feature, since for the Universe $r_*=r_c$ (see below).}
Depending on a particular gravitating source, the scale $r_*$ can be small enough so that the deviations can be potentially detectable by precision gravitational experiments  at much shorter (e. g., solar system) distances. 

 The model in question,  is a brane-world model of\cite{dgp}, which consist of a brane 
embedded in five-dimensional Minkowski space.  I shall use the notion of a brane in a very general sense
of a 4D sub-space on which the standard model particles live.  The action of the theory is
\beq
S~=~{M_P^2 \over 4r_c} ~ \int d^4x ~dy~ \sqrt{|g^{(5)}|}~ R_{5} 
 +~
{M_P^2 \over 2}~ \int d^4x ~ \sqrt{|g|}~ R~. 
\label{action}
\eeq
The first term is just an usual 5D Einstein action, with the five dimensional Plank mass given by
$M_*^3 \, = \, {M_P^2\over 4r_c}$.  $g_{\mu\nu}(x)$ is 
the induced metric on the brane, which in the approximation of the "rigid" brane is simply given by 
the value of the five-dimensional metric at the position of the brane. For instance,  if we locate the brane at  the origin of the fifth coordinate $y$, the induced metric 
takes the form  $g_{\mu\nu}(x) \, = \, g^{(5)}_{\mu\nu}(x, y =0)$. 
The 4D Einstein term on the
brane ($R$) plays the crucial role in generating 4D gravity on the brane at intermediate distances, 
despite the fact that  the space is actually
five-dimensional.  Here is how this effect comes about.  Consider a gravitating source localized 
on the brane, and let us ask what kind of gravitational field will it create on the brane. That is, what 
type of a Newtonian attractive force will it exert on the test bodies that are also localized on the brane? 
The Newtonian force is mediated by the virtual graviton exchange, which in our case
(after some gauge fixing) satisfy 
the following linearized equation
\begin{equation}
\label{linearized}
\left ( {1 \over r_c} (\nabla^2 \, - \,  \partial_y^2)\, + \, \delta(y)
\, \nabla^2
\right )\, h_{\mu\nu} \,  = \, -\, {1 \over M_P^2} \,\left \{ T_{\mu\nu} -{1\over 3}\eta_{\mu\nu} 
T^\alpha_\alpha   \right \}\delta(y)
+ \delta(y) \,\partial_\mu
\partial_\nu \,h^\alpha_\alpha\,,  
\end{equation}
where $T_{\mu\nu}$ is the brane-localized energy-momentum source.
The unusual thing about this graviton is that its propagation is governed by two kinetic terms. 
 The 4D kinetic term localized on the brane forces the graviton to propagate according 
to four-dimensional laws, which would result into the 4D $1/r^2$-Newton's force. 
The 5D kinetic term, however, allows graviton  to also propagate off the brane, and this term alone 
would of course result into a 5D Newton's force, which scales as  $1/r^3$.  In the presence of 
both kinetic terms
graviton "compromises", and the force law exhibits the crossover behavior.   For $r\ll r_c$ the potential is
$1/r$, and for $r\gg r_c$ it turns into $1/r^2$.

 The effect of generation of (almost) four-dimensional gravitational interaction at intermediate distances, 
due to {\it weakening} of  five-dimensional gravity on the brane, we shall call 
{\it shielding}\cite{DGKN}.   The essence of this effect is that  the large kinetic term on the brane makes
it hard for a graviton, with the extra momentum exceeding the critical value $1/r_c$,  to penetrate the brane.  Hence,  the number of gravitons that the brane-localized sources are exposed to
is much lower then the analogous number for the bulk sources.  This creates an effect of a weak 
4D-gravity as opposed to strong 5D one.  In the Newtonian approximation the effect 
can be best understood in terms of the four-dimensional
mode expansion. From this perspective a high-dimensional
graviton represents a continuum of four-dimensional  Kaluza-Klein (KK) states and
can be expanded in these states. This KK decomposition  can
schematically
be written as follows:
\begin{equation}
  h_{\mu\nu}(x,y)\, = \,\int_0^{\infty} dm\, h_{\mu\nu}^{(m)}(x)\,
 \sigma_m (y)\,,
\label{kkexp}
\end{equation}
where  $h_{\mu\nu}^{(m)}(x)$ are the four-dimensional spin-2 fields of mass
$m$ and   $\sigma_m(y)$ are their wave-function profiles
in extra dimension. The strength of the coupling of an individual mode of mass $m$
to a brane observer is given by the value of the wave-function at the
position of the brane, that is  $\sigma_m(0)$.  It is a very special 
$m$-dependence  of this function that makes our model different from simple
5D gravity. 

Gravitational force on the brane is mediated by
exchange of all the above modes.  An each KK-mode gives rise to
an Yukawa-type gravitational  potential of range $m^{-1}$, and the net result is
\begin{equation}
 V(r)\,\propto \,{1 \over M_*^3}\,
\int_0^\infty dm\, |\sigma_m(0)|^2
\,{e^{-rm} \over r}\,.
\label{kkpot}
\end{equation}
So far, this is a generic expression for any 5D gravity theory.  At the level of the Newtonian 
potential, all the difference between our theory and the ordinary 5D gravity is encoded 
in function $\sigma_m(0)$. In an ordinary  five-dimensional gravity we would have
$|\sigma_m(0)| \, = \, 1$,  meaning that all the KK-modes couple to brane-sources with an equal 
strength. As a result the integral (\ref{kkpot}) would sum up  into the usual 
5D potential  $\sim 1/M_*^3r^2$. In our case, because of shielding, the wave-functions of KK 
gravitons on the brane have the following form \cite{DGKN1}
\beq
|\sigma_m(0)|^2\, = \,{4 \over 4 + m^2r_c^2}\,,
\label{N2prof}
\eeq
which diminishes for $m \gg r_c$, according to the shielding effect.  
Given this form, it is a trivial exercise to see that the integral (\ref{kkpot}) exhibits an interpolating
behavior between the four and five dimensional regimes at the scale $r_c$.
 An equivalent way to interpret the above force is to think of it a being mediated by a four-dimensional 
resonance of the width $1/r_c$. 

\section{Implications for the Dark Energy.}
 Before going to the discussion of more fundamental issues, let me briefly discuss the cosmological implications of IR-modification.  Suppose we have an observer, that lives on the brane and knows nothing about the existence of extra dimensions. Based on cosmological observations this observer derives an effective cosmological equation for the
four-dimensional scale factor $a(t)$, of the effective 4D Friedmann-Robertson-Walker (FRW) metric.
The question is what would be the analog of the 4D Friedmann equation derived in such a way?
The answer turns out to be the following (see \cite{ddg} for details) 
\begin{equation}
\label{FRW}
H^2 \, \pm \, H/r_c \, = {8\pi G_N \over 3} \rho
\end{equation}
where $H\, = \, \dot{a} /a$ is the Hubble parameter, and $G_N$ is the four-dimensional Newton's
constant.  The $1/r_c$ correction comes from IR-modification of gravity, due to fifth dimension, and is negligible at early times. So for $H \gg 1/r_c$ the standard FRW cosmology is reproduced. However, for late
times modification is dramatic.  In particular, the cosmological expansion admits a "self-accelerating" 
branch with constant $H = 1/r_c$, without need of any matter source\cite{cd,ddg}. 
Hence, the large-distance modification of gravity may be a possible explanation for the late
time acceleration of the Universe, that is suggested by the current observations\cite{super}. 
Because of the specific nature of the transition, the upcoming precision cosmological studies
can potentially differentiate between the modified gravity and more conventional dark energy scenarios\cite{ddg,dt}. 

 It is expected that such a late time deSitter phase is a generic property of 
IR-modified gravity theories, since such modifications  should  result in non-linearities in $H$ in 
the modified Friedmann equation, and thus, in general could admit new solutions with constant $H$, even for $\rho = 0$. 

\section{Is it Alive?}

 Let us come back to the gravitational potential and ask the question whether such modification is ruled out. Consider a one-graviton exchange amplitude between the two brane-localized sources 
 $T_{\mu\nu}$ and  $T^{\prime}_{\alpha\beta}$ (the sign tilde denotes the quantities
which are Fourier transformed to momentum space)
\begin{equation}
{\cal A} \,
\,= \, -\frac{8\pi\,G_N}{q^2+ q/r_c}\left ( {\tilde
T}_{\mu\nu}\,-\,{1\over 3}\,
\eta_{\mu\nu} 
\, {\tilde T}^\beta_\beta   \right )\, {\tilde T}^{\prime\mu\nu}\,.
\label{dgp}
\end{equation}
where $q = \sqrt{q_{\mu}q^{\mu}}$ is a four momentum.   
The analogous amplitude in GR would have the form 
\begin{equation}
{\cal A}_0\,
=\,- \frac{8\pi\,G_N}{q^2}\left 
( {\tilde T}_{\mu\nu}\,-\,{1\over 2}\,
\eta_{\mu\nu} 
\,{\tilde T}^\beta_\beta   \right )\, {\tilde T}^{\prime\mu\nu}\,.
\label{gr}
\end{equation}
The two differences are immediately apparent. First, in our IR-modified theory, there is a correction to the
denominator that goes as $q/r_c$. This correction however is unimportant at momenta $q \gg 1/r_c$
(distances $r\, \ll \, r_c$). 
Another difference is the difference in the tensor structure of the second term, $1/3$ in our case versus 
$1/2$ in GR. This difference indicates that our graviton contains addition degrees of freedom, on top
of the usual two of the standard massless graviton.  In particular, there is an additional scalar attraction.  
So is this theory ruled out? 

Before answering this question, let me do a historic detour and discuss the
analogous issue in massive gravity.  In '71 van Dam, Veltman\cite{vDV} and Zakharov \cite{Zak} (vDVZ) suggested that graviton mass, no matter how small, was excluded by solar system observations. This conclusion was based on the existence of the discontinuity of the linearized graviton propagator in massive versus
massless theory. The propagator of the massive gravity, derived from the Pauli-Fierz action 
(the only ghost-free linearized action for a massive spin-2 particle) has the following form 
\beq
D^{(m)}_{\mu\nu ;\alpha\beta}(q)\,=\, 
\left(
{1\over 2} \,\tilde\eta_{\mu\alpha} \tilde \eta_{\nu\beta}+
{1\over 2} \, \tilde\eta_{\mu\beta}  \tilde\eta_{\nu\alpha}-   
{1\over 3} \, \tilde\eta_{\mu\nu} \tilde \eta_{\alpha\beta}\right)\frac{1}{
q^2+m_g^2-
i\epsilon}\,,
\label{5D}
\eeq  
where 
\begin{equation}
\tilde\eta_{\mu\nu}=\eta_{\mu\nu}+\frac{q_\mu q_\nu}{m_g^2}\,.
\end{equation}
Whereas, the massless graviton propagator is 
\beq
D^{(0)}_{\mu\nu ;\alpha\beta}(q)\,=\, \left(
{1\over 2} \,\eta_{\mu\alpha}  \eta_{\nu\beta}+
{1\over 2} \, \eta_{\mu\beta}  \eta_{\nu\alpha}-   
{1\over 2} \, \eta_{\mu\nu}  \eta_{\alpha\beta}\right)\frac{1}{
q^2-
i\epsilon}\,,         
\label{4D}
\eeq  
where only the momentum independent parts
of the tensor structure is kept. By a gauge  choice the momentum dependent
structures can be taken to be zero.
On the other hand, there is no such gauge freedom for the massive gravity.

Although the propagator of massive graviton contains terms  singular in $m_g$,  they  vanish when convoluted with conserved sources, and thus, are irrelevant in one-particle exchange. 
Hence, if the one-graviton exchange were a good approximation, the vDVZ discontinuity 
indeed would rule out the massive gravity.  Assumption of the one-graviton exchange dominance 
is therefore implicit in vDVZ analysis.  However, in {\it massive} theory the validity of  one-graviton approximation is  questionable for solar system distances,  as it was first argued by Vainshtein\cite{av} 
shortly after vDVZ works. 

 In\cite{ddgv} the issue was reconsidered in the following light. Imagine that we wish to 
study the motion of a planet in the massive theory. We can try to find the metric in form of the 
$G_N$-exansion. But in order to trust this expansion,  one has to be sure that the leading effect is indeed given by one-graviton exchange, that is by the contribution of order $G_N$. However, this is not always the
case, because the additional polarizations of massive graviton have couplings singular in $m_g$
which {\it do not necessarily}  vanish beyond the linearized approximation.  In fact one can directly check that the 
"subleading" diagrams proportional to $G_N^2$,
at solar system distances are about $10^{32}$-times the "leading"  ones proportional to
$G_N$! This invalidates vDVZ conclusion: although there is a discontinuity in the propagator, this discontinuity sheds no light on the phenomenological validity of the massive theory, since 
at solar system scales the linearized approximation for massive theory breaks down. 
Breakdown of the linearized gravitational approximation at the solar system distances 
may come as a surprise, however, we must remember that we are dealing with the
modified theory that includes {\it additional} degrees of freedom (additional polarizations of massive graviton), not present in the massless theory. 
As shown in \cite{ddgv} the breakdown of perturbation theory is due to strong coupling of the longitudinal polarizations of the massive graviton, due to which a trilinear vertex
exhibits $1/m_g^4$-type singularity (higher in $1/m_g$ singularities cancel).  The same conclusion
was reached in \cite{m2}, using the Stuckelberg type formalism.  

 Although the breakdown of the perturbation theory invalidates the phenomenological objections based
on vDVZ-discontinuity of the linearized graviton propagator, at this level the issue of the validity of the massive graviton theory remains inconclusive due to the lack of a fully generally-covariant 
ghost-free formulation of such a theory.\footnote{Nonlinear completions of Pauli-Fierz action also suffer from instabilities\cite{bd,gg}.}
 On the other hand, the model\cite{dgp} is generally-covariant and is free of ghosts, and the argument  can be taken through there.   
There too, the linearized propagator exhibits a vDVZ-type discontinuity in $r_c \rightarrow \infty$ limit, and there
too the additional polarizations of graviton have couplings singular in $1/r_c$, invalidating 
perturbation  theory at solar system scales\cite{ddgv}.  However, the correctly resummed solutions
can be found and are continuous. 

The effective brane-to-brane graviton propagator can be written as
\beq
D_{\mu\nu ;\alpha\beta}(q)\,=\, 
\left(
{1\over 2} \,\tilde\eta_{\mu\alpha} \tilde \eta_{\nu\beta}+
{1\over 2} \, \tilde\eta_{\mu\beta}  \tilde\eta_{\nu\alpha}-   
{1\over 3} \, \tilde\eta_{\mu\nu} \tilde \eta_{\alpha\beta}\right)\frac{1}{
q^2+q/r_c-
i\epsilon}\,,
\label{dgpp}
\eeq  
where 
\begin{equation}
\tilde\eta_{\mu\nu}=\eta_{\mu\nu}+\frac{q_\mu q_\nu}{q/r_c}\,.
\end{equation}
and we see that it shares some important features with massive gravity in the sense that the 
additional polarizations are  identical (one spin zero and two spin one), and the role of the
mass is played by $q/r_c$ term. This is not surprising since the 5D graviton from the point of view of the
4D theory looks like a resonance that can be spectrally expanded in continuum of massive spin-2
states (\ref{kkexp}), each of which has five polarizations.  It can be directly checked that the tree-graviton vertex exhibits the following singularity \cite{ddgv}
\begin{equation}
\label{strong}
 {q^3r_c^2 \over M_P}
\end{equation}
and hence the longitudinal polarizations become strongly coupled at the scale $q_s \, = \,(M_P/r_c^2)^{{1 \over 3}}$. (This was also found in \cite{prl} in the language of
\cite{m2}). 
What is the physical meaning of this scale?  The breakdown of the linearized approximation and existence of the "strong coupling" signals that $G_N$ is not any more a good expansion parameter, and series have to be re-summed.  If the resummation is possible and the answer does not explicitly contain the "strong coupling" scale, then this scale is simply an artifact of the incorrect
perturbative expansion in powers of $G_N$.  
The breakdown of the perturbative expansion (at least at the solar system scales) and the existence of the re-summation are the key points for avoiding the  vDVZ conclusion\cite{ddgv}. 

At the tree-level there indeed is a complete resummation for different cases of 
interest\cite{ddgv, lue, andrei}.
For instance, Schwarzchild solution can be found in terms of $1/r_c$-expansions and exhibits a complete continuity.  This solution has the following  form \cite{andrei}
\begin{eqnarray}
&&\nu(r) \,=  \,-{r_g\over r}+{\cal O}\left({1 \over r_cr}\sqrt{r_g
r^3}\right)\,,\qquad
\lambda(r) \,= \,{r_g\over r}+{\cal O}\left({1 \over r_cr}\sqrt{r_g
r^3}\right)\,,\nonumber\\[1mm]
\label{sw}
\end{eqnarray}
Here $r_g$ is the gravitational radius of the body, and the functions $\nu(r)$ and $\lambda(r)$ parameterize the spherically-symmetric metric in the usual way
$g_{00}\, = \, {\rm e}^{\nu(r)}, \, g_{rr} \, = \, {\rm e}^{\lambda(r)}$. 
As we see, any reference to the scale $q_s$ has disappeared from the resummed solution.
Which indicates that the "strong coupling" scale could indeed be an artifact of the
perturbative expansion in terms of $G_N$. 

We could ask the question about the effect of this  strong coupling behavior on the
loop corrections.  It is obvious that since the perturbative expansion in $G_N$ breaks down already
at the tree-level, it will also break down in the loops. The question is whether a similar resummation 
should exist at the loop level, or the tree-level resummation is some sort of an accident? 
According to one possible believe \cite{prl}, the loops should ruin resummation, 
but this is far from being obvious. 
So far, the only supporting argument for such a statement is that the perturbation theory breaks down in 
loops, which is obvious, as it breaks down already at the tree-level due to exactly the same
non-linear interaction\cite{ddgv}.  
Unfortunately, the only way to compute loops in non-renormalizable theory is to know its 
ultraviolet (UV)  completion.
Simply cutting off the loop divergences at the presumed scale is illegitimate, and moreover adds nothing to our knowledge unless we have some independent information about the UV scale,  and at least can do matching of scales (as e.g.,  in the chiral Lagrangians). 
UV-completion, such as embedding in string theory (see \cite{strings1,strings2}) potentially could answer this 
question, but at the level of the effective IR-theory the effect of the loop corrections is unclear. 
Phenomenologically, the most important question is not whether the perturbation theory breaks down
(which is the fact),  but whether the UV-completed theory (in which one can properly account for loops) 
affects the validity of the resummed solution obtained in the IR-theory.   We shall discuss this issue
in more details later on an explicit example that indicates that the UV-completion is not affecting 
the resummed solution. 

Coming back to the
solution (\ref{sw}), we notice that it does exhibit some interesting behavior. 
Let us focus on the form of the corrections to the ordinary Schwarzchild metric.  This corrections are
suppressed
by powers of $r/r_c$ as it should be the case by continuity. However they are also enhanced 
by powers of the ration $r/r_g$. This enhancement  has a clear physical  meaning.  Indeed, as we have seen
the continuity is restored by non-linearities that are important near the gravitating sources. 
This is where the effect of additional polarization diminishes.  For the light sources, however, the 
non-linearities are unimportant and corrections due to additional polarizations become 
significant.  As a result such deviations could be potentially observed near the light sources. To briefly summarize, there are the following important scales that govern the metric of gravitating 
spherically-symmetric bodies.  These are $r_g$,  $r_c$ and a new scale 
\begin{equation}
\label{rstar}
r_* \, = \, (r_c \sqrt{r_g})^{{2\over 3}} 
\end{equation}
Regimes of gravity alternate as follows.  For $r\ll r_*$ gravity is essentially Einsteinian, with small corrections given by (\ref{sw}). For $r_* \ll r \ll r_c$ gravity is still $1/r$ 
but is  of a scalar-tensor type. Finally for $r \gg r_c$ gravity becomes 5D  $1/r^2$-type. 

 \section{Meaning of the "Strong Coupling" Scales.}

 I wish now to discuss what are the physical scales of the theory.  
For example,  when dealing with the perturbative expansion in series of $G_N$ we
have encountered a "strong coupling" type behavior at the scale
$q_s$.  What is the meaning of this scale, and what are its effects on observations? As mentioned above, one possibility is 
that this scale is simply an artifact of the perturbative expansion in $G_N$, and should disappear
when the perturbative series are correctly resummed.  This possibility is strongly supported  by the existence of 
the fully-resummed tree-level solutions.  If the loop expansion would admit similar resummation, then
the scale $q_s$ would be unphysical. 

  There of course remains an option that the tree-level resummation is some sort of an accident, and there is no analog effect at the loop-level. In such a case the scale $q_s$ may be physical, 
meaning that 
some physical degrees of freedom, not accounted in the effective low energy theory, have to be introduced around this scale.  The question we want to ask now is how strongly these new degrees of freedom modify the resummed tree-level solutions of the effective theory above the scale $q_s$. In the other words, how much the solutions
obtained in low energy effective theory differ from the analogous solutions obtained in 
UV-completed theory above the scales $q_s$?  If the tree-level resummation is an accident, then
the usual expectation is that the two solutions should differ dramatically above the scale $q_s$. 
However, the situation is more subtle. We wish to argue that 
in many cases the form of the tree-level solution is unaffected by the UV-completion, including loop
corrections. 

 We shall now discuss an explicit toy model with the above property. 
It is a model with a non-Abelian  Goldstone field, whose effective low energy description
shares many similarities with the gravity case, including a strong-coupling behavior above the certain scale.  However, the explicit UV-completion of this model demonstrates that
classical solutions on the brane are unaffected by the UV-completing physics even above the
naive "strong coupling" scale. 

 Consider the following low energy action
\begin{equation}
\label{sigma}
S \, = \, M_*^3 \, \int \, d^5x \, \partial^{A} \,u^a \partial_{A} \, u^a \,  +  \, 
 M^2\, \int\, d^4x \,\partial^{\mu} u^a\partial_{\mu} u^a \,
\end{equation}
Where $A$ is the five-dimensional index, and $M$ and $M_*$ are some mass parameters. Below we shall be interested in the limit
$M_* \ll M$.
$u^a,~ a=1,...4$ is an $O(4)$-vector obtained by an arbitrary coordinate-dependent $O(4)$-rotation
of the unit vector $(1,0,0,0)$.  This rotation can be parameterized by the three Goldstone angles
that we shall call $\phi (x), \theta(x)$ and  $\chi(x)$.  The vector $u$ then can be written in the following form 
\begin{equation}
\label{u}
u \, =\, (c_{\phi}s_{\theta}s_{\chi}, \, s_{\phi}s_{\theta}s_{\chi}, \, c_{\theta}s_{\chi},\, c_{\chi})
\end{equation}
where $c$ and $s$ stand for Cos and Sin functions of the respective angles. 
Theory is obviously invariant under a global $O(4)$-rotation under which the Goldstone fields
transform non-linearly, whereas the vacuum is only $O(3)$-invariant.   
The model (\ref{sigma}) is a close analog of the model of infrared-modified gravity given 
by the action (\ref{action}).  
Goldstone fields play the role analogous to graviton.  Just like gravitons
Goldstones are also universally coupled,  and their non-linear interactions also exhibit strong coupling behavior.   Because of the interplay of the
two kinetic terms, the above model also exhibits the shielding phenomenon, due to  which
the linearized interactions on the brane interpolate between 4D and 5D regimes
at the crossover scale $r_c \, = \, M^2/M_*^3$.  

 The effective theory has at least one "obvious" strong coupling scale, $M_*$.  Indeed, the interactions 
of the 5D Goldstone fields  are cut-off by this scale. So naively any classical solution on the brane, 
losses validity above the energy $M_*$.  Let us see if this is indeed the case.  Consider one such solution, which occurs for a spherically-symmetric brane of radius $R$.  To build it up, let us first restrict our 
attention to the 4D world-volume of the brane,  ignoring the fifth dimension. The world-volume theory
is a theory on a three-sphere, with the metric 
 $ds^2 \, = \, dt^2 - R^2[d\hat{\chi}^{2}  + s_{\hat{\chi}}^2
(d\hat{\theta}^2 \, + \, s_{\hat{\theta}}^2 \, d\hat{\phi}^{2})]$, where $\hat{\chi}, \, \hat{\theta}, \,\hat{\phi}$
are the usual angular variables.  Since the vacuum manifold is also a three-sphere,  on such a space there is a topologically non-trivial solution,  the "texture" \cite{texture}, which is obtained by mapping 
$S_3 \rightarrow S_3$. Texture is obtained from (\ref{u}) by assuming the following 
coordinate dependence of the Goldstone fields $\chi=\hat{\chi}, \, \theta=\hat{\theta}, \,\phi=\hat{\phi}$. 
The energy density of the texture comes purely from the Goldstone gradient energy and scales as
\begin{equation}
\label{e4texture }
E \sim  M^2/R^2
\end{equation}
Hence, in truly four-dimensional theory given by the second term in (\ref{sigma}), the solution would
be valid all the way to the curvature radius $R \sim M^{-1}$.  We may expect that the extension of the
Goldstone model into the 5D space with lower cut-off $M_*$, should jeopardize this validity above the 
scale $M_*$.   Indeed, let us first note that the solution can be easily embedded in 5D space, if 
 $\hat{\chi}, \, \hat{\theta}, \,\hat{\phi}$ are identified with spherical coordinates in 5D space
and we place the brane at $\rho = R$ (where $\rho$ is the radial coordinate in 5D).
The resulting configuration has the  form of the
hedgehog configuration of the vector $u$
\begin{equation}
\label{utext}
u_{text} \, =\, (c_{\hat{\phi}}s_{\hat{\theta}}s_{\hat{\chi}}, \, s_{\hat{\phi}}s_{\hat{\theta}}s_{\hat{\chi}}, \, c_{\hat{\theta}}s_{\hat{\chi}},\, c_{\hat{\chi}})
\end{equation}
In an effective theory (\ref{sigma}) this solution breaks down at distances $\rho \ll M_*$, where the 
five-dimensional angular gradients blow-up.
Thus, if the brane has a radius $R \gg 1/M_*$ the 4D texture solution is a good approximation
on the brane, but  should break down for the radius $1/M \ll R \ll 1/M_*$.  
Let us now show that this naive expectation is not supported in the UV-completed theory. Although, the
solution gets modified in the bulk, the
world-volume configuration remains a valid solution all the way to the scale of UV-completed theory
which can be arbitrarily higher than any strong-coupling scale of the effective IR theory. 

 The obvious UV completion of the above effective theory is given by the Goldstone model 
with a spontaneously broken  $O(4)$-symmetry by the Higgs field $\Phi^a$ transforming as a four-component vector.  The action for $\Phi$ we choose in the 
following way
\begin{equation}
\label{uvcomp}
S\, = \, \int \, d^5x \, \left ( \partial_A\Phi^a\partial^A\Phi^a \,   -  \, {1 \over 2M} (\Phi^a\Phi^a \, -\, M_*^3)^2
\right ) \, - \, \int \, d^4x  \, {1 \over 2M^2 } \, \left (\Phi^a\Phi^a \, - \, M^3\right )^2 
\end{equation}
The cutoff of the full UV-completed theory is 
$M$.
 This theory describes a spontaneously  broken $O(4)$-symmetry, but the scales of  breaking on the brane and in the bulk are very different. Scale of breaking  in the five-dimensional bulk is $M_*$, whereas, 
as we shall demonstrate in a moment, the scale of breaking in the brane world-volume is
$\sim M$.
We expect that there will be all possible high-dimensional operators suppressed by powers
of the scale $M$ generated from the quantum loops. These, however, 
cannot affect any of our conclusions as long as we keep the hierarchy $M_* \ll M$. 
This addresses the issue of the loop corrections 
in UV-completed theory. 
The difference from the previously  discussed sigma model is that the absolute value of
$\Phi$ can fluctuate, or in the other words the Goldstones are supplemented by the Higgs degree of freedom
\begin{equation}
\label{higgs}
\Phi \, = f(x) u(x)
\end{equation}
The mass of the Higgs excitation in the bulk is $m_{Higgs}^2 \sim  M_*^3/M$. 
 Hence, the effective low energy description in terms of the Goldstone degrees of freedom
in the bulk breaks down at the scale $M_{Higgs}$. 
Despite this fact the non-linear solution of the full UV-completed theory on the brane 
is well described by the solution (\ref{utext}) of the sigma model all the way to the scale $M$. 
As a result, for the brane 
world-volume observers the classical solutions  of the Goldstone theory remain valid up to 
the energies of order $M$.  Before convincing ourselves in this  validity, 
let us show that in the lowest energy state $\Phi$  indeed develops a large expectation value 
on the brane.  It is easiest to prove this in the limit $M_* \rightarrow 0$, in which case $\Phi$ 
assumes a zero expectation value in the bulk. The fact that $\Phi$ wants to condense on the brane can be seen by examining the linearized equation for small perturbations about the $\Phi = 0$ 
solution in the brane background, which has the following form (due to $O(4)$-symmetry it is enough to consider a single component of $\Phi$)   
\begin{equation}
\label{inst}
\left (\nabla^2 \, - \, \partial_y^2 \, - \delta(y) M  \right ) \Phi \, = \, 0 
\end{equation}
It is obvious that this equation has a  normalizable exponentially growing (imaginary frequency) tachyonic  eigen-mode, localized on the brane
\begin{equation}
\label{tachyoin}
\Phi \, = \, e^{{1\over 2} Mt} e^{-{1\over 2}|y|M}
\end{equation}
This instability signals that $\Phi$ condenses on the brane and develops a large expectation value
$\sim M$,  as both the tachyonic mass as well as stabilizing potential are set by a single mass scale $M$. 
 Hence, the expectation value on the brane is
$\sim M$ and in the bulk is $M_*$.  Then the low energy action for the
Goldstone fields is similar to (\ref{sigma}). 
Let us see what is the effect of this UV-completion on the texture solution on the brane 
at energies $\gg M_*$. 

For the spherically symmetric brane of radius $R$, the texture solution is embedded in 5D in form of the
hedgehog  configuration of the scalar field, which in the five dimensional spherical coordinates can be
written as 
\begin{equation}
\label{embtexture}
\Phi \, = \, f(\rho) \, u_{text}
\end{equation}
where the angular dependence of $u_{text}$ is as given in (\ref{utext}).  The  difference from the analogous solution in the Goldstone model, is that the absolute value of the vector $\Phi^a$ (given by function
($f(\rho)$))  can now respond to the
phase gradients, and vanish in the regions in which the phase-gradients become more costly than the 
Higgs energy.  The phase gradient energy density is  of order 
\begin{equation}
\label{gradiente}
E_{grad} \, \sim \, {f(\rho)^2 \over \rho^2}  \, 
\end{equation}
This has to be compared with bulk and brane Higgs energy densities.  
For distances $\rho \ll \sqrt{M/M_*^3}$ the bulk Higgs energy density becomes less costly than the gradient energy, and $O(4)$ symmetry gets restored. Hence, in the bulk $f(\rho)$ vanishes inside a four-dimensional sphere of radius $\rho \ll M_{Higgs}^{-1}$. This scale, however, cannot affect the validity of the non-linear 
brane world-volume solution.  Even for the branes with the size $R << 1/M_*$ this solution remains intact.  This is obvious, since on the world-volume the Higgs energy 
is more costly than the gradient energy all the way to the scales $\rho \sim 1/M$. Hence, the 
expectation value of $\Phi$ on the brane remains non-zero as long as the 
brane radius 
$R \gg  1/M$.  Thus, even for the brane sizes $1/M \ll R \ll 1/M_*$ the solution of the IR-theory stays in tact. 

 To interpolate between the different regimes,   we can first take $R\gg 1/M_{Higgs}$ and 
gradually decrease it to the 
values $R \ll 1/M_*$.  As long as $R \gg 1/M_{Higgs}$ the 5D-gradients of the Goldstone phases in the immediate neighborhood of the brane are small compared to the bulk Higgs energy .
Then the 5D  Higgs potential  wins and  the Higgs field stays in the vacuum both on the brane as well as  in the bulk.  
Hence solution (\ref{embtexture}) of the UV-completed theory coincides with the solution of 
IR-theory both on the brane as well as in the bulk. 

When the brane shrinks to the radius $1/M\ll R \ll 1/M_{Higgs}$, 
the 5D-Higgs potential
becomes sub-dominant to the 5D-gradients of the phase, and can no  longer support the non-zero
expectation value of $\Phi$ outside the brane.  In the other words, the brane enters inside the
core of the hedgehog solution, where symmetry is restored.  However, a four-dimensional observer on the brane
continues to see the spontaneously-broken $O(4)$-symmetry at scale $M$, because  the phase-gradients are much less
costly than the world-volume Higgs energy.  So texture solution on the brane is unaffected 
by the UV-regulating physics. 
For example, to unwind the texture on the world-volume, we have to make 
$\Phi$ vanish there, which is more costly  then the gradient energy. Hence $\Phi$ stays in the 
vacuum inside the brane, and texture solution stays in tact. 
 For the four-dimensional observer on the
brane the situation looks identical to what it would look in 4D space with spontaneously broken $O(4)$ symmetry at scale $M$.  Hence, there is no reference to the scales  $\sqrt{M/M_*^3}$ or 
$M_*$ as far as the solution on the brane is 
concerned, even above the eneregies of order $M_*$! 
 
 In the other words the strongly coupled bulk physics got "shielded" from the brane observer by the
huge brane-localized kinetic term.  Although  the above example is not gravity, it shows the 
generic point that UV-completions act "softly" on the brane-solutions in shielded theories. Hence, 
in this class of theories, IR-solutions can still be used for confronting observations.  

\section{Anomalous Perihelion Precession.}

 The considered theory predicts the slight modification of  the gravitational
potential of a massive body at the observed distances according to (\ref{sw}). This modification can be observable in
the experiments that are sensitive to anomalous perihelion precession of planets\cite{dgz}. 
Following the analysis of \cite{dgz}, let $\epsilon$ be the
fractional change of the gravitational potential
\begin{equation}
\epsilon \equiv {\delta \Psi \over \Psi },
\end{equation}
where $\Psi=-GM/r$ is the Newtonian potential. The anomalous perihelion
precession (the perihelion advance per orbit due to gravity
modification) is
\begin{equation}
\delta \phi =\pi r (r^2(r^{-1}\epsilon )')',
\end{equation}
where $'\equiv d/dr$.

Let us apply this to the model of \cite{dgp}.
In this theory, \cite{andrei}
\begin{equation}
\epsilon \, = \, - \, \sqrt{2}r_c^{-1}r_g^{-1/2}r^{3/2}.
\end{equation}
The numerical coefficient deserves some clarification. The above coefficient was derived in
\cite{andrei} on Minkowski background. However, non-linearities created by cosmological expansion can further correct the coefficient. One would expect these corrections to scale as
powers of $r_cH$, where $H$ is the observed value of the Hubble parameter. On the accelerated branch, as it's obvious from  (\ref{FRW}),  $H \, \sim \, 1/r_c$ and thus,  
one would expect the corrections to be of order one. 
We will restrict ourselves to order of magnitude estimate, but the sign may be important if the effect is found, since according to\cite{ls} it could give information about the cosmological branch.
We get
\begin{equation}
\delta \phi =(3\pi /4)\epsilon.
\end{equation}
Numerically, the gravitational radius of the Earth is $r_g=0.886$cm, the
Earth-Moon distance is $r=3.84\times 10^{10}$cm, the gravity
modification parameter that gives the observed acceleration without dark
energy $r_c=6$ Gpc. We get the theoretical precession
\begin{equation}
\delta \phi = 1. 4 \times 10^{-12}.
\end{equation}
This is to be compared to the accuracy of the precession measurement by
the lunar laser ranging.
Today the accuracy is $\sigma _\phi
=4\times 10^{-12}$ and no anomalous precession is detected at this
accuracy, in the near future a tenfold improvement of the accuracy is
expected\cite{er}.
Then a detection of gravity modification predicted by \cite{dgp} will be possible.

 Although we have focused on a particular model 
of large-distance gravity modification, we expect that the property that the corrections to the ordinary gravity near the weak sources should set in at relatively shorter distances
should be shared by the whole class of such theories.   The reason is that, as argued above, 
the theories that modify gravity in far IR propagate additional polarizations, and hence, at the linearized level give the 
results different from GR (or else they have to contain ghosts, and are not viable).
So in order to be compatible with observations, in these theories the naive
perturbative expansions in $G_N$ must break down at least at the solar system distances, and strong
coupling of additional polarizations precisely accomplishes this role\cite{ddgv}.  The correctly re-summed solutions  should exhibit continuity in $1/r_c$.  From this reasoning it follows that in any theory
that passes the solar system tests the resummed solution near the gravitating sources must have the property  that  the relative deviations from the GR are enhanced by powers of  $r/r_g$. 
Indeed, for small $r_g$
(the weak sources) the linearized gravity, which is by order one different from the Einsteinian counterpart for the same source,  must become  a better approximation at relatively shorter distances than for the heavy sources. Hence the continuity demands that at a fixed distance $r$ the corrections should increase with decreasing  $r_g$.   Thus, the fractional corrections to the Einsteinian gravitational potential are expected to be of the form
\begin{equation}
\epsilon \sim r_c^{-1-\gamma }r_g^{-1/2}r^{3/2+\gamma },
\end{equation}
where $\gamma$ is model-dependent.  Notice that the fact that the cosmological 
evolution of the Universe should be dramatically modified when the horizon becomes 
comparable to $r_c$, directly follows  from the above expression. Indeed, treating  the Universe as the source with the gravitational radius
of $r_g \sim r_c$, the fractional correction  $\epsilon$ becomes order one. 
The anomalous lunar perihelion precession is
\begin{equation}
\delta \phi \sim r_c^{-1-\gamma }r_g^{-1/2}r^{3/2+\gamma }.
\end{equation}
\label{precgamma}
Observations of accuracy $\sigma_ \phi$ can therefore test gravity
theories with
\begin{equation}
r_c < r \left( {r\over \sigma _\phi ^2r_g}\right) ^{1\over
2(1+\gamma )}.
\end{equation}
One can speculate that in the absence of additional scales 
the gravity theories that produce self-acceleration, without vacuum energy should have
 $r_c\sim$few Gpc. Then the lunar precession accuracy of $\sigma _\phi \sim 10^{-12}$ will tests theÊ $\gamma =0$ theory\cite{dgp}.

\section{Discussion and Outlook.}

Possibility of large-distance modification of gravity is a fundamental question, motivated by 
the Dark Energy\cite{ddg,dt,ian,inverse} and the Cosmological Constant\cite{dgs,eva,addg} problems. 
In this lecture we have addressed the issue of viability and observable impact of such a modification. 
Although for the illustrative purposes we have focused mostly on a concrete model (\ref{action}), our
conclusions are very general and apply to large class of theories.  In particular,  this class should include
any generally-covariant  theory
in which the graviton propagator admits a sensible spectral representation in terms of  positive norm
mass- and spin-eigenstates.  
 
 The following generic features emerged.  IR-modified gravity theories propagate additional 
degrees of freedom (additional polarizations of graviton).  Due to these addition states,  at the the linearized level the predictions of modified theories are by order one different from the predictions of GR.   Hence, the only way the new theories could be compatible with the experiment is if the naive perturbation theory in terms of $G_N$-expansion breaks down, at least at solar system distances.  It is interesting that in the discussed model this is 
precisely what happens, due to the strong coupling of the additional polarizations of the graviton\cite{ddgv}.  
Appearance  of the strong coupling in the naive perturbative expansion  is the only reason why
these  theories are not immediately ruled out.  At the classical level there is a full resummation 
and non-linear solutions exhibit continuity in $1/r_c$.  The naive strong coupling scale disappears 
from  these solutions.  Of course, there is a question of loops, which is impossible to answer without
knowing the UV-completion.  Doing loop expansion in terms of $G_N$ and simply cutting off the loop divergences at the suspected scale, 
knowing that the perturbation theory in the very same expansion parameter breaks down already at the tree-level, adds nothing to our understanding of the situation, unless we have some information
about the UV-completion. 

 However, even without knowing the precise form of the UV-completion, one could envisage the two options.  One not unlikely possibility is that the tree-level resummation is not an accident, and indicates that an analogous  resummation  should take place in the loops.  In such a case the scale $q_s$
should be disregarded as the artifact of the incorrect perturbation theory.

  It is certainly possible that the resummation is an accident, and does not go beyond the three-level.  
 Then the perturbative strong coupling probably indicates that there are new degrees of 
freedom at the scale $q_s$ and these should be integrated in. In such a case the theory above $q_s$ will be  described by the UV-completed theory that will include these new degrees of freedom. 
The crucial question then is how different are the fully-resummed non-linear solutions of
IR-theory, which exhibit no pathology above the scale $q_s$, in comparison with the analogous 
solutions of the UV-completed theory? 
  Although, one may naively expect that the difference between the two must become dramatic above the scale $q_s$, the suggestive toy examples exhibit  the opposite behavior. The solution on the brane is 
essentially unaffected by the UV-completion.  We suspect that this may be a property of the theories in which four-dimensional behavior on the brane is obtained by shielding. This connection will be
discussed elsewhere. 

 Due to new degrees of freedom, the theories with IR-modified gravity are very constrained and can be potentially tested  not only by the precision cosmology, but also by shorter scale gravitational measurements, and in some models (that predict an especially low fundamental scale of gravitational interaction) may be probed by the upcoming collider experiments \cite{DGKN}.   

\vspace{0.5cm}   

{\bf Note Added}
\vspace{0.1cm} \\

Since this talk was given there have been some interesting developments, and for brevity, I shall only
comment on the ones that are relevant for the issue of the "strong coupling" discussed here. 
Modifying the model \cite{dgp}, Gabadadze and Shifman \cite{gs04} (see also \cite{pr04})  have constructed an example that even in  the straightforward perturbative expansion remains weakly coupled till the  scale $M_*$.  This is an interesting existence proof, 
perhaps complementary to the approach 
outlined in this talk, according to which
the breakdown of the perturbative $G_N$-expansion is the way for making
IR-modified gravity (with additional polarizations) compatible with observations.
Luckily this is accomplished by the singular in 
$1/r_c$ "strong coupling" of extra polarizations, as discussed in \cite{ddgv}. 
The key to the issue is in correct resummation which eliminates the dependence 
of the physical solutions on the strong coupling scale.

\vspace{0.5cm}   

{\bf Acknowledgments}
\vspace{0.1cm} \\

It is the great pleasure to thank  the organizers of the Nobel Symposium for the hospitality at this very exciting meeting.  I would like to thank G.~Gabadadze and A.~Gruzinov for comments. This work is supported in part  by David and Lucile  Packard Foundation Fellowship for  Science and Engineering, and by NSF grant  PHY-0070787.

\end{document}